\documentstyle[prl,aps,multicol,epsf]{revtex}
\begin{document} 
\draft 
\title{Single particle excitations of the Kondo lattice}

\author{Robert Eder, Oana Stoica, and George A. Sawatzky}
\address{Department of Applied and Solid State Physics, 
University of Groningen,\\ 
9747 AG Groningen, The Netherlands}
\date{\today}
\maketitle

\begin{abstract} 
We present a simple theory for the description of the single particle
excitations of the Kondo lattice model. We derive an `effective
Hamiltonian' which describes the coherent propagation of single 
particle-like fluctuations on a strong coupling groundstate. Even for 
$f$-electrons which are replaced by Kondo-spins, the resulting spectral 
function obeys the Luttinger theorem {\em including} the $f$-electrons, 
and our calculation reproduces the complicated evolution of the
spectral function with electron density seen in numerical studies.
\end{abstract} 
\pacs{71.27.+a, 71.28.+d, 75.20.Hr} 
\begin{multicols}{2}

The theoretical description of the Kondo lattice 
remains an unsolved problem of solid state physics, yet
the solution of this problem is crucial
for understanding the anomalous prperties 
of $f$-electron metals\cite{Fulde}.
The simplest model which incorporates the essential physics
is the Kondo lattice model
\begin{eqnarray}
H &=& 
\sum_{\bbox{k},\sigma} \epsilon(\bbox{k})c_{\bbox{k},\sigma}^\dagger 
c_{\bbox{k},\sigma} 
- V \sum_{i,\sigma} (c_{i,\sigma}^\dagger f_{i,\sigma} + H.c.)
\nonumber \\
&-& \epsilon_f \sum_{i,\sigma} n_{i,\sigma}
+ U\sum_{i} n_{i,\uparrow} n_{i,\downarrow}.
\label{kondo1}
\end{eqnarray}
Here we consider the `minimal' model, where each
unit cell contains two orbitals, one of them for the
uncorrelated conduction electrons
the other for the strongly correlated $f$-electrons. Then, 
$c_{i,\sigma}^\dagger$  ($f_{i,\sigma}^\dagger$)
creates a conduction electron ($f$-electron) in cell $i$, 
$n_{i,\sigma}$$=$$f_{i,\sigma}^\dagger f_{i,\sigma}$,
and $\epsilon(\bbox{k})$$=$$\sum_{i,j}
exp(i\bbox{k} \cdot(\bbox{R}_i - \bbox{R}_j))\; t_{i,j}$
is the Fourier transform of the inter-cell
hopping integral $t_{i,j}$ for $c$ electrons.
For simplicity we consider only the symmetric case, $U=2\epsilon_f$.
For $U$$\gg$$V$ (\ref{kondo1}) can be reduced to
\begin{equation}
H_{sc} =
\sum_{\bbox{k},\sigma} \epsilon(\bbox{k})c_{\bbox{k},\sigma}^\dagger 
c_{\bbox{k},\sigma} 
+ J \sum_i \bbox{S}_{i,c} \cdot \bbox{S}_{i,f}
\label{kondo2}
\end{equation}
where $\bbox{S}_{i,c}$ $(\bbox{S}_{i,f})$ denotes the
spin operator for conduction electrons ($f$-electrons)
in cell $i$ and $J=4V^2/\epsilon_f$.
One problem which by many is believed to be at the heart of the
solution is the way in which the more or less
localized $f$-electrons,
which are replaced by mere spin degrees of freedom in the strong coupling 
version (\ref{kondo2}) ,
participate in the formation of the Fermi surface.
De Haas-van Alphen experiments on heavy Fermion metals\cite{Tailefer} 
as well as computer simulations of
Kondo lattice models\cite{Tsutsui,Moukuri1} suggest that despite their
`frozen' charge degrees of freedom the $f$-electrons participate in the
Fermi surface volume as if they were uncorrelated. 
The limiting cases $V$$=$$0$, $J$$=$$0$, which obviously
do not allow for participation of the $f$ electrons in the Fermi surface,
therefore represent singular points, so that a perturbation expansion
in the (small) parameters $V$ or $J$ may not be expected to give
meaningful results. Rather, the interaction between $f$-spins and conduction
electrons must be incorporated in a non-perturbative
way, in a similar manner as the single-impurity Kondo effect\cite{Wilson}.
It is the purpose of the present manuscript to present
a minimum effort theory for the Kondo lattice which is based
on this requirement and
shows how the nominal participation of the localized electrons in the
Fermi surface can be
understood even in the complete absence of true hybridization.
We describe the system by an `effective Hamiltonian' for the
Fermion-like charge fluctuations on top of a strong coupling ground state,
and show that this treatment leads to remarkable agreement
with the scenario inferred  from numerical simulations.\\
As a starting point, we take the case of half-filling (i.e.
two electrons/unit cell, corresponding to the `Kondo insulator') 
and vanishing inter-cell hopping integrals $t_{i,j}$.
In this limit, the lattice ground state is simply the product
of $N$ single-cell ground states, each of them being 
%%%%%%%%%%%%%%%%%%%%%%%%%%%%%%%%%%%%%%%%%%%%%%%%%%%%%%%%%%%%
\begin{figure}
\epsfxsize=8.1cm
\vspace{-6.5cm}
\hspace{-1.0cm}\epsffile{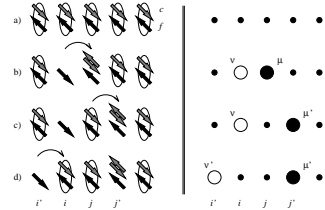}
\vspace{-0.5cm}
\narrowtext
\caption[]{Charge fluctuations and their propagation
(left panel) and their representation in terms of `model Fermions'
(right panel).}
\label{fig1} 
\end{figure}
%%%%%%%%%%%%%%%%%%%%%%%%%%%%%%%%%%%%%%%%%%%%%%%%%%%%%%%%%%%%
\noindent 
a two-electron
singlet state (see the state (a) in Figure \ref{fig1}).
In the following we consider this product state as a kind of
`vaccuum'. This state clearly is a total singlet, and has no
magnetic order; it thus is appropriate to discuss the paramagnetic
phase of primary interest. Switching on the $t_{i,j}$ 
then produces `charge fluctuations'
on this vacuum: an electron can jump from a cell
$i$ to another cell $j$, leaving
the cell $i$ in a single electron eigenstate with number $\nu$,
and the cell $j$ in a three electron eigenstate with number $\mu$
(see state (b) in Figure \ref{fig1}). 
We note that the Hamiltonian (\ref{kondo1}) allows for two such
eigen states of a single cell, the strong coupling version (\ref{kondo2})
only for one.
In further steps, these charge fluctuations can
propagate: an electron from the three-fold occupied
cell $j$ can hop to another neighbor $j'$ leaving
cell $j$ with two electrons and $j'$ in a three electron state
(see Figure \ref{fig1}c ) or, alternatively,
an electron can jump from
another neighbor $i'$ into cell $i$, leaving $i$ in
a two electron state, $i'$ in a single electron state.
When there is only one $c$ orbital/unit cell,
the single cell states of both
$1$ or $3$ electrons are spin dublets, i.e. these states
have the spin quantum numbers of ordinary electrons.
As our basic approximation we now restrict the Hilbert space such
that whenever there are $2$ electrons in one cell, they are
in the singlet ground state. This restriction implies that the
propagating charge fluctuations do not `leave a trace' of excited
cells, i.e. their motion under this constraint becomes completely coherent.
We can then interpret the first step in Figure \ref{fig1}
as a pair creation process, where two `book keeping Fermions'
are created on nearest neighbors in the vacuum state,
and the subsequent steps in
Figure \ref{fig1} as  a propagation of these Fermions.
More precisely, if cell number $i$ is in
the $\nu^{th}$ single electron eigenstate with $z$-spin $\sigma$
we interpret this as presence of a hole-like
Fermion, created by $a_{\nu,\sigma}^\dagger$, whereas
the cell being in the $\mu^{th}$ three electron eigenstate 
with $z$-spin $\sigma$ is modelled by the presence of an electron-like
Fermion, created by $b_{\mu,\sigma}^\dagger$.
Within our restricted Hilbert space
the dynamics of the model then is decribed by the Hamiltonian
$H_{eff}$$=$${\cal P} H {\cal P}$ where
\begin{eqnarray}
H = \sum_{i,\sigma} (\; \Delta\;
a_{i,\sigma}^\dagger a_{i,\sigma} +
\bar{\Delta}\; b_{i,\sigma}^\dagger b_{i,\sigma}\;) +
\sum_{i,j,\sigma} [\;(\;
V_{i,j}\; b_{j,\sigma}^\dagger a_{i,\bar{\sigma}}^\dagger
\nonumber\\
+ V'_{i,j}\; a_{j,\sigma}^\dagger a_{i,\sigma}
+ V''_{i,j,}\; b_{j,\sigma}^\dagger b_{i,\sigma}\;)
+ H.c.\;] .
\label{eff}
\end{eqnarray}
and ${\cal P}$ projects onto the subspace
of states where no site is occupied by more than one
Fermion. This kinematic constraint reflects the
fact that the state of a given cell must be unique.
(for the sake of brevity we have
suppressed the state indices $\mu,\nu$ in (\ref{eff})).
The procedere is very much analogous to
spin wave theory, with the sole difference that we are considering
Fermionic charge fluctuations rather than Bosonic spin fluctuations.
Due to the product nature of the basis states, the
various matrix elements $V$, $V'$ and $V''$ in (\ref{eff}) are easily
expressed in terms of the electron removal
and addition matrix elements of a single cell,
$m_{PES,\nu} = \langle \Psi_\nu^{(1)} | c_\sigma | \Psi_0^{(2)} \rangle$
and
$m_{IPES,\mu} = \langle \Psi_\mu^{(3)} | c_\sigma^\dagger 
| \Psi_0^{(2)} \rangle$. For example
$V_{(i,\nu),(j,\mu)}$$=$$ t_{i,j} \;m_{PES,\nu}\; m_{IPES,\mu}$
(here $|\Psi_\mu^{(n)}\rangle$ denotes the $\mu^{th}$ eigenstate
of a cell with $n$ electrons).
The `on site energies' are defined
as $\Delta_\nu$$=$$E_\nu - E_0$.\\
In analogy with linear spin wave theory we now relax
the constraint enforced by ${\cal P}$, 
whereupon the Hamiltonian (\ref{eff}) is readily solved
by Bogoliubov transformation,
$\gamma_{\bbox{k},\lambda,\sigma}^\dagger$$
=$$\sum_\nu u_{\bbox{k},\lambda,\sigma}^{(\nu)}
a_{\nu,-\bbox{k},\bar{ \sigma}}
+ \sum_\mu u_{\bbox{k},\lambda,\sigma}^{(\mu)}
b_{\mu,\bbox{k},\sigma}^\dagger$.
A possibility to treat the constraint in a more rigorous
fashion would be to apply complete Gutzwiller projection to
the wave functions obtained from (\ref{eff}); anticipating that the
main effect is a renormalization
of the matrix elements, we may expect that a calculation without
the constraint at least qualitatively reproduces the correct
one. Our main justification for this approximation, however, is the good
agreement with exact cluster results as discussed below.
As mentioned above,
the Hamiltonian (\ref{eff}) may be thought of as describing the
coherent propagation of
single particle-like fluctuations `on top of' the strong coupling
`vacuum' state. 
One may expect that there will also be e.g. spin-like
fluctuations, which would correspond to a cell being 
occupied by two electrons in a triplet state. The creation
of these spin excitations by the propagating charge fluctuations
would be described by terms of the form
$\;a_{i,\tau}^\dagger\vec{S}_{j,}^\dagger\cdot
 \vec{\sigma}_{\tau,\tau'} a_{j,\tau'}$,
where $\vec{S}_{j}^\dagger$ is a bosonic 
spin-triplet operator. Such terms could be treated
e.g. within the `rainbow diagram' approximation\cite{SchmidtRink},
but in the present manuscript we restrict ourselves to
the coherent motion.\\
For the strong coupling limit (\ref{kondo2}) the calculation becomes
particularly simple: one finds 
$m_{PES}$$=$$m_{IPES}$$=1/\sqrt{2}$
whence $V_{i,j}$$=$$-V'_{i,j}$$=$$V''_{i,j}$$=$$t_{i,j}/2$,
$\Delta$$=$$\bar{\Delta}$$=$$3J/4$, so that the dispersion relation
reads
\begin{equation}
E_\pm(\bbox{k}) = (1/2)\;[\; \epsilon(\bbox{k}) \pm
\sqrt{ \Delta^2 + \epsilon(\bbox{k})^2}\;].
\label{scdisp}
\end{equation}
Formally, this is equivalent to the hybridization
of a dispersionless `effective' $f$-level in the band center
with a free electron band with dispersion $\epsilon(\bbox{k})$,
the strength of the `nominal' mixing element being $3J/4$.
It should be noted, however, that the 
resulting energy gap of $\Delta$ does not arise from
the formation of a bonding and antibonding combination of
$c$-like and $f$-like Bloch states, as in the hybridization model;
rather, this gap originates from the energy cost to break
two intra-cell singlets in the course of a charge fluctuation.
This gap therefore is of a similar nature as the energy gap in a 
superconductor.\\
Having computed the eigenvalue spectrum, we proceed to the
computation of the single particle spectral function.
This requires to resolve the `ordinary' electron operators
in terms of the model Fermions. Taking into account
our basic assumption, namely that a single cell 
with $2$
electrons can only be in its ground state, we can expand
e.g. the electron annihilation operator as
\begin{equation}
c_{i,\sigma} = \sum_\nu m_{PES,\nu} a_{i,\nu,\bar{\sigma}}^\dagger
 + \sum_\mu m_{IPES,\mu}^* b_{i,\mu,\sigma}.
\label{exp}
\end{equation}
Then, Figure \ref{fig2}a shows the single particle spectral function
for a $1D$ chain of the full
Kondo lattice (\ref{kondo1}) at 
half-filling (see e.g. Ref.\cite{Tsutsui} for a precise definition
of this quantity). 
Particle-hole symmetry implies
that the Fermi energy is zero in this case.
It is obvious at first sight, that
our calculation is in excellent
agreement with the results of exact diagonalization\cite{Tsutsui}.
To begin with, unlike any band theory approach,
our calculation gives the correct number of
`bands': the nearly 
dispersionless upper and lower Hubbard bands
with almost pure $f$ character, and the two `hybridization bands'
which resemble the result for the strong coupling limit,
(\ref{scdisp}).
The hybridization bands change their character from
high intensity $c$-like to weak intensity $f$-like near
$\pi/2$, i.e. the Fermi momentum of the unhybridized $d$ electrons.
Note that while the dispersion of these bands could be modelled by
band theory, the sharp drop of spectral weight at $\pi/2$ could not.
In the Kondo limit, where the strength of the ` effective hybridization'
is small, there are extended regions of flat 
%%%%%%%%%%%%%%%%%%%%%%%%%%%%%%%%%%%%%%%%%%%%%%%%%%%%%%%%%%%%
\begin{figure}
\epsfxsize=8cm
\vspace{-0.5cm}
\hspace{0.5cm}\epsffile{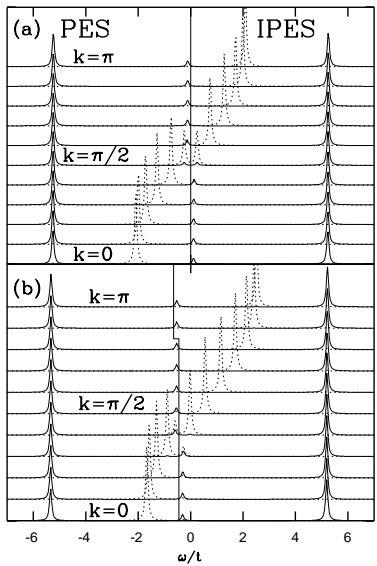}
\vspace{-0cm}
\narrowtext
\caption[]{Single particle spectral function for the $1D$
Kondo lattice with $\epsilon(k)$$=$$-2\cos(k)$,
$U$$=$$10$ and $V$$=0.5$.
The electron density is $2.0$/unit cell in (a)
(i.e. half-filling)  and $1.6$/unit cell
in (b). Full lines (dashed lines) correspond to 
$f$-like ($c$-like) spectral weight. 
Peaks to the right (left) of the vertical line
correspond to electron creation (annihilation).}
\label{fig2} 
\end{figure}
%%%%%%%%%%%%%%%%%%%%%%%%%%%%%%%%%%%%%%%%%%%%%%%%%%%%%%%%%%%%
\noindent
(i.e. `heavy') bands. 
Quite obviously the basic idea of our approach, namely
to broaden the ionization and affinity states of a single cell
into bands works well as far as the dispersion of energy
and spectral weight is concerned. The present approximation does not
incorporate the infrared divergences which occur in the
impurity problem, and therefore does not reproduce the
parameter dependences of the
low energy scales\cite{Yoshida}. For example the lowering of
the total energy/site as compared to the noninteracting case is
$\sim \Delta^2 ln(W/\Delta)$
when calculated from (\ref{scdisp}), and the $f$-like
spectral weight $Z_f$ of the
`heavy bands' has the same functional dependence on
the parameter values as in the single cell problem, i.e.
$Z$$\sim$$(V/\epsilon_f)^2$. \\
We proceed to the doped case and first consider the 
expression for the electron number operator.
We give explicit expressions only for the strong coupling limit
(\ref{kondo2}), but all considerations for the full Kondo lattice
are completely analogous.
In the `vacuum state', the number of electrons is $2N$ and the
presence of an $a$-Fermion ($b$-Fermion)
decreases (increases) the electron number by $1$, so that the
electron number operator should be simply
\begin{equation}
N_e=\sum_{\bbox{k},\sigma} (\;
a_{\bbox{k},\sigma} a_{\bbox{k},\sigma}^\dagger + 
b_{\bbox{k},\sigma}^\dagger b_{\bbox{k},\sigma}
\;) = \sum_{\bbox{k},\mu,\sigma}
\gamma_{\bbox{k},\mu,\sigma}^\dagger \gamma_{\bbox{k},\mu,\sigma}.
\label{count}
\end{equation}
This operator counts both localized and conduction electrons,
so that reducing the electron density
$\rho_e$ below $2$ will give a  Fermi edge in the lower
hybridization band which satisfies
a `nominal' Luttinger theorem, i.e. including the $f$ -electrons.
The physical origin, however, is the fact that
we have a density $2-\rho_e$ of holes in the insulating
`singlet background'.
This results in a `hole pocket' of volume
$(1-\rho_e/2)$, which however is
nominally equivalent to a Luttinger Fermi surface {\em including} 
the $f$-electrons. There is, however, an extra complication:
the electron number should also equal the integrated
PES weight, which is given by the expectation value of
\begin{equation}
N_e' = 
N + \sum_{\bbox{k},\sigma} 
 c_{\bbox{k},\sigma}^\dagger c_{\bbox{k},\sigma}.
\label{wcount}
\end{equation}
The first term on the r.h.s. is the contribution of the
$f$-electrons' lower Hubbard band.
Using (\ref{exp}),
which now reads $c_{\bbox{k},\sigma}$$=$$(1/\sqrt{2})
(a_{\bbox{k},\sigma}^\dagger + b_{\bbox{k},\sigma})$,
this becomes
\[
N + \frac{1}{2}\sum_{\bbox{k},\sigma} 
(a_{\bbox{k},\sigma} a_{\bbox{k},\sigma}^\dagger +
b_{\bbox{k},\sigma}^\dagger b_{\bbox{k},\sigma}
+a_{\bbox{k},\sigma}^\dagger b_{\bbox{k},\sigma}^\dagger
+ b_{\bbox{k},\sigma} a_{\bbox{k},\sigma} ).
\]
Counting the electrons in real
space on one hand and integrating the spectral weight
on the other hand thus results in different expressions
for the electron number. This can hardly be a surprise, since
in a strongly correlated electron system spectral weight
and band structure are decoupled to a large extent.
As an example let us consider the extreme Kondo limit, and
assume that we are gradually reducing the electron density from the
`Kondo insulator value' of $2$.
The chemical potential
corresponding to the real-space electron count (\ref{count})
then will cut more and more into the `heavy' band, to
produce the nominal Luttinger Fermi surface volume.
Since the spectral weight per $\bbox{k}$-point is much smaller
than $1$ for the `heavy' band, however,
the amount of spectral weight which crosses from PES to IPES
as the the Fermi momentum changes
can never be consistent with the electron number.
A simple rigid band picture therefore must fail
to maintain the consistency of electron number and
integrated spectral weight.
To cope with this problem, we choose the simplest possible solution and 
enforce the consistency of `ordinary' 
electron count and spectral weight integration
by adding both expressions, (\ref{count}) and (\ref{wcount}),
for the electron number to the Hamiltonian,
each of them multiplied by a separate Lagrange multiplier:
$H \rightarrow H + \mu N_e + \lambda N_e'$.
The notion of `two chemical potentials' may seem
awkward at first sight, but as we will show now, this approach 
results in a remarkable consistency with the numerical results.
For the strong coupling limit (\ref{kondo2}) this substitution
gives the same dispersion relation as (\ref{scdisp}),
however with the replacements
$E(\bbox{k})$$\rightarrow$$E(\bbox{k})$$-$$\mu$ and
$\epsilon(\bbox{k})$$\rightarrow$$\epsilon(\bbox{k})$$-$$\lambda$.
The effect of these modifications can be seen in Figure \ref{fig2}b,
which shows the spectral function for the Kondo lattice away
from half-filling.
To begin with, the $f$-like upper and lower hubbard band remain
unaffected by the change in electron concentration.
The Lagrange multiplier
$\mu$ acts like a standard chemical potential,
which cuts into the `heavy' band and, as discussed above,
produces a Fermi 
%%%%%%%%%%%%%%%%%%%%%%%%%%%%%%%%%%%%%%%%%%%%%%%%%%%%%%%%%%%%
\begin{figure}
\epsfxsize=8cm
\vspace{-0.5cm}
\hspace{0.5cm}\epsffile{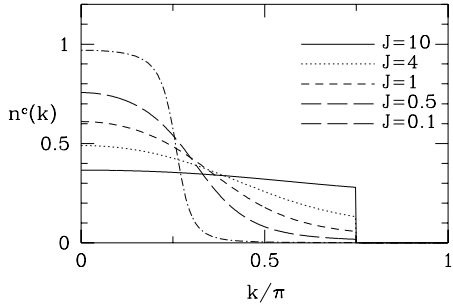}
\vspace{-3.5cm}
\narrowtext
\caption[]{Momentum distribution for the
strong coupling limit for different values of $J$, 
$\rho_c=0.5$.}
\label{fig3} 
\end{figure}
%%%%%%%%%%%%%%%%%%%%%%%%%%%%%%%%%%%%%%%%%%%%%%%%%%%%%%%%%%%%
\noindent 
surface consistent with the nominal Luttinger theorem.
On the other hand, the Lagrange multiplier for the
spectral weight, $\lambda$, modifies the dispersion relation
far from the true Fermi momentum: the occupied part of the
$c$-like band structure shortens, the unoccupied part grows.
In fact, the change of the $c$-like spectral weight
alone is very reminiscent of what is expected for
free (i.e. unhybridized) conduction electrons.
This can be understood by carrying on
the formal analogy of (\ref{scdisp}) with a hybridization
gap picture: $\lambda$ obviously plays the role of the `on-site
energy' of the effective $f$ level.
Next, in the Kondo limit
$J\ll t$ the momentum distribution for the conduction
electrons, $n_c(\bbox{k})$ has the limiting behaviour
$n_c(\bbox{k}) \approx 1$ for $(\epsilon(\bbox{k})-\lambda) \ll J$,
$n_c(\bbox{k}) \approx 0$ for $(\epsilon(\bbox{k})-\lambda) \gg J$,
and $n_c(\bbox{k})=1/2$ for $\epsilon(\bbox{k})=\lambda$.
The constant energy surface $\epsilon(\bbox{k})=\lambda$ therefore represents
a kind of `pseudo Fermi surface', where
$n_c(\bbox{k})$ drops from nearly $1$ to nearly$0$
over a distance of order $\Delta/v_F$ (with $v_F$ the Fermi
velocity of the conduction electrons).
Since we have the sum rule $(2/N)\sum_{\bbox{k}} n_c(\bbox{k})$$=$$\rho_c$,
the density of conduction electrons, it follows immediately
that $\lambda \approx E_F^0$,
the Fermi energy for the unhybridized conduction electrons
{\em without} counting the $f$-spins.
In other words, the band structure of the Kondo lattice
away from half-filling
is equivalent to an effective $f$-level, which is pinned near
the `frozen core' Fermi energy for conduction electrons
of density $\rho_c$, and mixes with a matrix element
of strength $\Delta$. This non-rigid band like
change of the spectral function upon doping is precisely what
was seen the cluster diagonalization results
of Tsutsui {\em et al.}\cite{Tsutsui}. Next, 
Figure \ref{fig3} shows  the change of
$n_c(\bbox{k})$ with $J$ for a $1D$ chain
of the strong coupling limit (\ref{kondo2}).
For very large $J$ (which is an unphysical limiting case)
$n_c(\bbox{k})$ is nearly constant and drops to $0$ at
$k_F=3\pi/4$, the Fermi momentum for
{\em hybridized} conduction and $f$ electrons.
As $J$ is reduced, the true
Fermi surface discontinuity shrinks more and more,
and the `pseudo Fermi surface' at $k_F^0$$=$$\pi/4$, the
`frozen core' Fermi momentum, 
starts to develop. Again, this behaviour of $n_c(\bbox{k})$
is in complete agreement with numerical results\cite{Ueda,Moukuri2}.\\
In summary,we have outlined a simple approximate theory for the
single particle excitations of the Kondo lattice model. 
It may be viewed as the construction of an effective Hamiltonian
describing Fermionic fluctuations on a strong coupling
vacuum state. While the calculation
requires a number of rather strong approximations, and consequently 
may fail
to correctly reproduce the extreme low energy energy scales of the
problem, we believe that it also offers a number of advantages:
despite its extraordinary simplicity the obtained results
compare favourably with the secnario obtained in numerical simulations, 
and include some of the known constraints, such as the
Luttinger Fermi surface and the paramagnetic
singlet nature of the ground state; one
may therefore hope hope that our approach captures the essential physics
of the problem.
The calculation moreover
is independent of dimensionality or lattice geometry, and hence may
be complementary to more sophisticated methods available 
in $1D$\cite{Tsvelik}.
Moreover, there are a number of obvious refinements, such as
a more rigorous treatment of the kinematic constraint
or the inclusion of spin-like fluctuations. \\
This work was supported by Nederlands Stichting voor Fundamenteel
Onderzoek der Materie (FOM) and Stichting Scheikundig Onderzoek
Nederland (SON). Financial support of R. E. by the European
Commuinity and of O. S. by the Soros Foundation is most gratefully
acknowledged.

\end{multicols}
\end{document}